\documentclass[pra,aps,a4paper,twocolumn,repreprint,superscriptaddress,longbibliography]{revtex4}

\usepackage{subcaption}
\usepackage{physics}
\usepackage{amssymb}
\usepackage{amsmath}
\usepackage{amsfonts}
\usepackage{hyperref}
\usepackage{graphicx}
\usepackage[dvipsnames]{xcolor}
\usepackage{bm}
\usepackage{chemformula}
\usepackage{caption}
\usepackage[justification=justified,format=plain]{caption}

\DeclareMathOperator{\sign}{sign}
\DeclareMathOperator{\Pf}{Pf}

\def\bn{\mathbf{n}}

\def\ve{\varepsilon}

\definecolor{mygold}{rgb}{0.93,0.69,0.13}

\definecolor{mypurple}{rgb}{0.49,0.18,0.56}

\newcommand{\eps}{\varepsilon}

\def\be{\begin{equation}}
\def\ee{\end{equation}}

\begin{document}

\title{Quantum critical dynamical response of the twisted Kitaev spin chain}

\author{Uliana E. Khodaeva}
\affiliation{Technical University of Munich, TUM School of Natural Sciences, Physics Department, 85748 Garching, Germany}

\author{Dmitry L. Kovrizhin}
\affiliation{LPTM, CY Cergy Paris Universite, UMR CNRS 8089, Pontoise 95032 Cergy-Pontoise Cedex, France}

\author{Johannes Knolle}
\affiliation{Technical University of Munich, TUM School of Natural Sciences, Physics Department, 85748 Garching, Germany}
\affiliation{Munich Center for Quantum Science and Technology (MCQST), Schellingstr. 4, 80799 M{\"u}nchen, Germany}
\affiliation{Blackett Laboratory, Imperial College London, London SW7 2AZ, United Kingdom}

\begin{abstract}
The dynamical structure factor of the transverse field Ising model (TFIM) shows universal power-law divergence at its quantum critical point, signatures of which have been arguably observed in inelastic neutron scattering studies of quantum spin chain materials, for example $\ch{CoNb2O6}$. However, it has been recently suggested that its microscopic description is better captured in terms of a twisted Kitaev spin chain (TKSC) with bond-anisotropic couplings. Here, we present exact results for the dynamical structure factor of the TKSC across its quantum critical point, analyzing both the universal low-frequency response and the non-universal high-energy features. In addition, we explore extensions of the model including broken glide symmetry as well as the case of random, and  incommensurate magnetic fields. Notably, in the latter case the fermionic excitations exhibit a localization-delocalization transition, which is manifest in the dynamical response as a distinct signature at finite frequency. We discuss the relevance of these features for the observation of quantum critical response in experiments.  
\end{abstract}

\date{\today} 

\maketitle

\section{Introduction}
Quantum phase transitions (QPT) are driven by quantum fluctuations and are typically induced by a variation in magnetic field, pressure, or chemical composition~\cite{sachdev2011}. Unlike classical phase transitions, which are governed by thermal fluctuations, QPTs occur at strictly zero temperature and are characterized by the interplay of competing ground states, giving rise to rich critical phenomena~\cite{Vojta_2003}. QPTs are manifest in the dynamical response as universal scaling behavior at low energies, but often exhibit intriguing non-universal features at higher energies. One-dimensional quantum spin systems, such as TFIM, have emerged as ideal platforms for studying QPTs both theoretically and experimentally~\cite{Pfeuty:1970qrn, Coldea_2010}.

In this context, the quantum spin material $\ch{CoNb2O6}$ has garnered significant interest for its potential realization of the TFIM and in the studies of quantum critical behavior~\cite{Coldea_2010,kinross2014evolution,liang2015heat,steinberg2019nmr}. However, experiments revealed significant deviations from the behavior expected in an ideal TFIM~\cite{HEID1995123, PhysRevB.63.024415,fava2020glide}. Morris et al.~\cite{Morris_2021} showed that the experimental data is better described by a different model which they dubbed the twisted Kitaev chain (TKSC). The TKSC is a representative of a general class of compass models and can be understood as an interpolation between the Ising model and the Kitaev chain~\cite{PhysRevB.89.104425}. 

Inelastic neutron scattering is one of the most widely used experimental probes to investigate quantum magnetism, making theoretical calculations of the dynamical structure factor $S(k,\omega)$ critically important~\cite{Sturm+1993+233+242}. Notably, the results of these calculations should not only capture the universal low-frequency response, but also predict non-universal high-frequency features. Although universal low-frequency properties offer fundamental insights into the nature of quantum phase transitions~\cite{sachdev2011}, they are often challenging to measure experimentally due to finite resolution, system size limitations, and disorder effects. In contrast, high-frequency features, which are sensitive to the microscopic details of the system, are frequently more accessible in experiments and can serve as valuable probes even in disordered samples.

The dynamical structure factor of the twisted Kitaev spin chain (TKSC) can be calculated exactly for an arbitrary frequency and wave vector using Jordan-Wigner transformation and the Pfaffian technique~\cite{mccoy2014two}. This method has been used to calculate the full $S(k,\omega)$ of the TFIM~\cite{PhysRevB.56.11659}, and employed in many studies of the TFIM with randomness~\cite{Young_1996, PhysRevB.56.11691}, as well as in the studies of quantum critical lines and fans in the TFIM with three-spin interactions~\cite{PhysRevB.108.155143, PhysRevB.107.045124}. Although similar results can also be obtained using DMRG~\cite{Laurell_2023}, the Pfaffian approach offers exact results for benchmarking and is capable of handling much larger system sizes and finer frequency resolution. Apart from the paradigmatic TFIM almost no other exact results of quantum critical response functions are known for lattice models.

In this paper, we calculate the dynamical structure factor of the TKSC across its quantum critical point and investigate not only the universal low-frequency response but also the non-universal high-energy features. We do not restrict our calculations to the conventional TKSC, but generalize this model in several different ways. First, we introduce the breaking of the glide symmetry~\cite{fava2020glide} and examine emerging high-frequency features in the dynamical structure factor. Second, we study the TKSC in a random transverse magnetic field. This modification is of particular interest since quantum phase transitions in disordered systems remain poorly understood~\cite{vojta2019disorder}. Although the random transverse field Ising model was extensively studied~\cite{PhysRevB.56.11691, Young_1996, Fisher_1998, PhysRevB.51.6411}, most famously via the strong disorder renormalization group by Fisher~\cite{PhysRevLett.69.534}, the dynamical response in the disordered critical case remains an open problem, and is beyond the standard Harris-type arguments~\cite{vojta2019disorder}. The interplay between conventional critical singularities and the "Griffiths-McCoy" singularities inherent to disordered systems gives rise to diverse phenomena, which can be explored using the method described in this paper. Third, we investigate quasiperiodic systems~\cite{PhysRevLett.48.1043,aubry_andre} that lie between translationally invariant and disordered limits. A qualitatively new feature of these systems is the possibility for localization-delocalization phase transitions in one dimension (1D), as well as the appearance of mobility edges~\cite{PhysRevB.96.085119}. For the case of the TFIM the rich phase diagram of the quasiperiodic 'Ising spin glass' has been studied~\cite{PhysRevX.7.031061}, but the distinct signatures in the dynamical structure factor have remained unexplored.

The paper is organized as follows. In Section II, we introduce the model and outline the corresponding Jordan-Wigner transformation. In Section III, we apply the Pfaffian technique to calculate the structure factor. In Sections IV and V, we present and discuss the results for the conventional TKSC, and the TKSC with broken glide symmetry. In Section VI, we investigate the TKSC in a random transverse field. Section VII is dedicated to exploring the physics of quasiperiodic case. Finally, we conclude in Section VIII with the discussion of our main results.

\section{The Model}
\begin{figure}
    \raggedright
    \captionsetup[subfigure]{labelformat=empty}
     \begin{subfigure}[b]{0.5\textwidth}
    \raggedright
        \includegraphics[width=6.8cm]{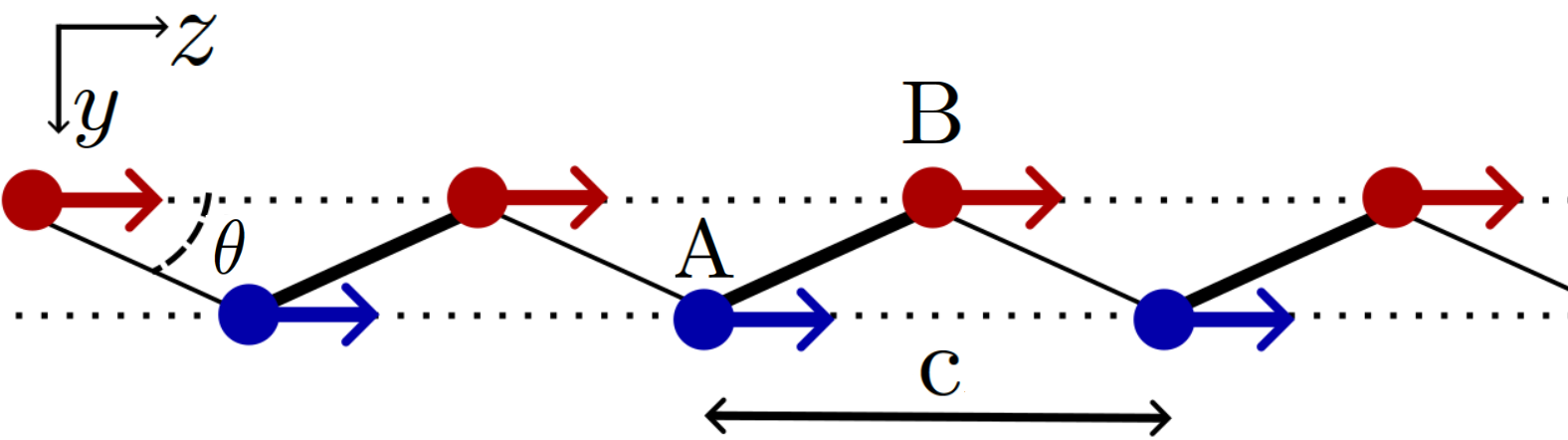} 
       \caption{\large{a)}} \label{Fig1:scheme}
    \end{subfigure}
     \vspace{1cm}
    \begin{subfigure}[b]{0.5\textwidth}
          \raggedright
        \includegraphics[width=8.0 cm]{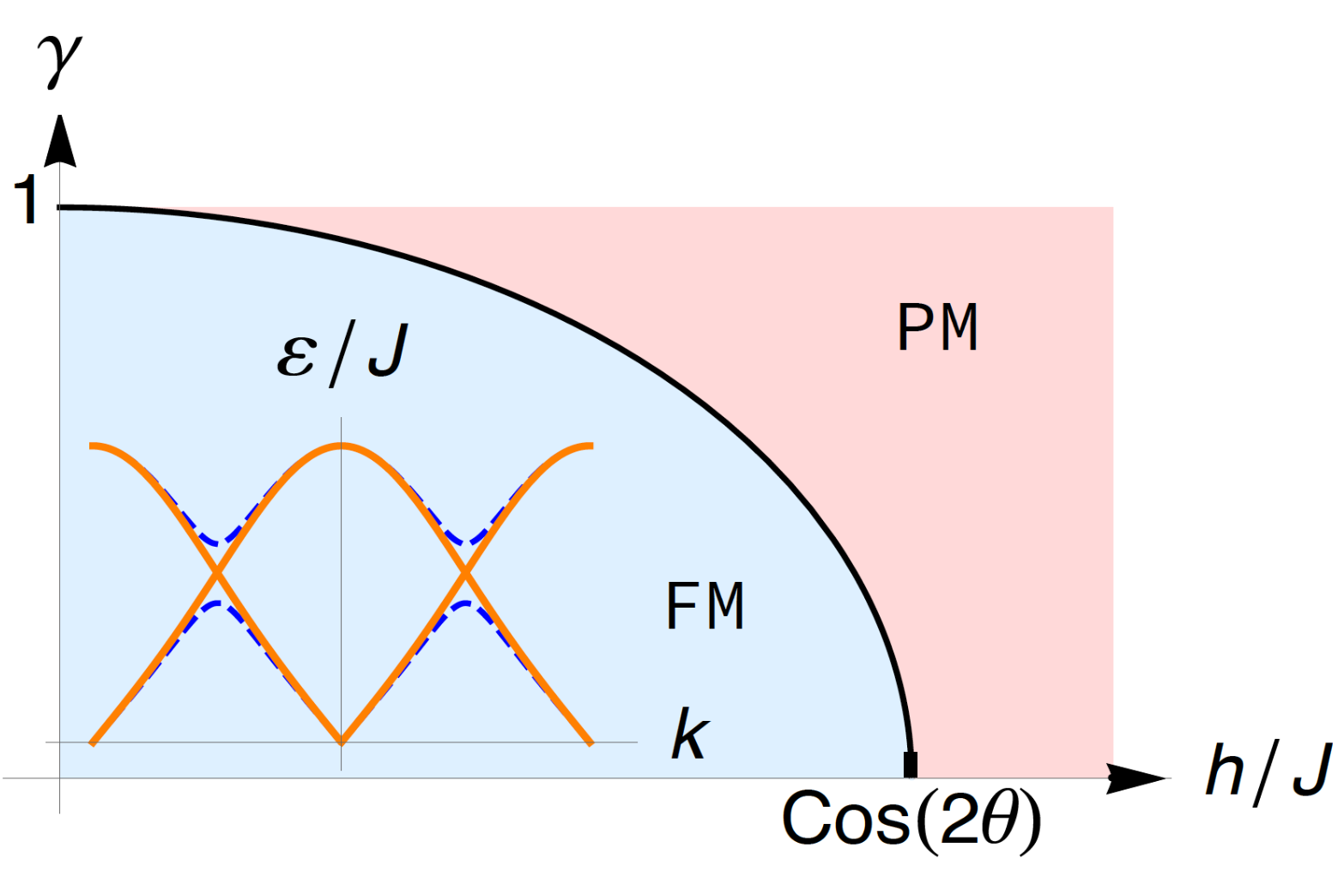} 
        \caption{\large b)} \label{Fig1:phase_diag}
    \end{subfigure}
    \caption{\textbf{Twisted Kitaev spin chain with broken glide symmetry.} \subref{Fig1:scheme}) Schematic picture of a twisted Kitaev chain, where the twist is characterized by an angle $\theta$. The difference in line thickness reflects the variation of bond strength in the case of broken glide symmetry. A unit cell consists of two sites, so the lattice has two sublattices, A (blue) and B (red). \subref{Fig1:phase_diag}) The phase diagram for  $\theta = \pi/10$. The blue domain corresponds to the ferromagnetic phase, and the red represents the paramagnetic phase. Note that the position of the quantum critical point depends on the value of $\gamma$. The inset shows the spectrum of the TKSC for $\theta = \pi/10$ at the critical point with broken glide symmetry (the dashed blue line, $\gamma = 0.2$) and without glide symmetry breaking (the solid orange line, $\gamma = 0$).}\label{tkc_scheme}
\end{figure}
We study TKSC~\cite{Morris_2021} in a transverse magnetic field and discuss a number of its solvable generalizations. The TKSC is a representative of a general class of compass models~\cite{Kliment_Kugel_1982},~\cite{RevModPhys.87.1} and is described by the following Hamiltonian (see Fig.\ref{tkc_scheme}):
\be
\hat{H} = -J\sum_{i = 1}^{L'}\left(\hat{\sigma}^{\hat{n}_1}_{2i - 1}\hat{\sigma}^{\hat{n}_1}_{2i} + \hat{\sigma}^{\hat{n}_2}_{2i}\hat{\sigma}^{\hat{n}_2}_{2i+1}\right) - h \sum_{i = 1}^L \hat{\sigma}^x_i, \label{tkc_hamilt_abstr}
\ee
where $J > 0$ is the ferromagnetic exchange parameter, $L' = L /2$ is the number of unit cells, each containing two sites, and $\hat{\sigma}^{\hat{n}_i}_j = \hat{n}_i\cdot\hat{\sigma}_j$ is the projection of the pseudospin Pauli operator vector on site $j$ onto the bond-dependent Ising direction $\hat{n}_i$. The name of this model derives from the fact that it interpolates between the Ising chain, $\hat{n}_1 \cdot \hat{n}_2 = 1$, and a one-dimensional section of the Kitaev model,  $\hat{n}_1 \cdot \hat{n}_2 = 0$.

For the sake of generality, we introduce a Hamiltonian describing a model with broken glide symmetry~\cite{Morris_2021,fava2020glide}, which interchanges $\bn_1$ and $\bn_2$:
\be
\begin{split}
\hat{H} = &-J\sum_{i = 1}^{L'}\left[\left( 1 + \gamma \right)\hat{\sigma}^{\hat{n}_1}_{2i - 1}\hat{\sigma}^{\hat{n}_1}_{2i} + \left( 1 - \gamma \right)\hat{\sigma}^{\hat{n}_2}_{2i}\hat{\sigma}^{\hat{n}_2}_{2i+1}\right]\\
&- h \sum_{i = 1}^L \hat{\sigma}^x_i, \label{tkc_hamilt_abstr_broken}
\end{split}
\ee
with the parameter $\gamma$ representing the strength of the symmetry breaking. Physically, this parameter reflects the magnitude of the distortion arising from, for example, a Peierls instability, which causes one bond in the unit cell to become stronger than the other. The standard TKSC model [Eq.\eqref{tkc_hamilt_abstr}] corresponds to the limit of $\gamma = 0$.

We choose the coordinate system in such a way that the Ising axes lie in the $yz$-plane and the $\hat{z}$-direction bisects the angle between $\hat{n}_1$ and $\hat{n}_2$, which is tantamount to the following parametrization: $\hat{n}_1 = (0, \sin\theta, \cos\theta)^T$ and $\hat{n}_2 = (0, \sin\theta, -\cos\theta)^T$. In these axes, the Hamiltonian above [Eq.\eqref{tkc_hamilt_abstr_broken}] can be expressed as
\be
\begin{split}
\hat{H} &= -J\sum_{i = 1}^{L}\left(1 + (-1)^i\gamma\right)\left[\sin^2\theta\,\hat{\sigma}^{y}_{i}\hat{\sigma}^{y}_{i+1} + \cos^2\theta\,\hat{\sigma}^{z}_{i}\hat{\sigma}^{z}_{i+1} \right.\\
&\left. + (-1)^i\sin\theta\cos\theta\left(\hat{\sigma}^{y}_{i}\hat{\sigma}^{z}_{i+1} + \hat{\sigma}^{z}_{i}\hat{\sigma}^{y}_{i+1}\right)\right] - h \sum_{i=1}^{L} \hat{\sigma}^x_i.
\end{split} \label{tkc_hamilt_spins}
\ee

When the Ising axes are non-collinear, $\hat{n}_1 \cdot \hat{n}_2 \neq 1$, i.e., $\theta \neq 0$, the system becomes frustrated. Generically, the frustration is minimized by forming a ferromagnet polarized in a direction that makes the smallest equal angle between the Ising axes~\cite{PhysRevB.89.104425} -- in terms of our parameterization, the $\hat{z}$-direction. An exception to this behavior is the Kitaev chain ($\hat{n}_1 \cdot \hat{n}_2 = 0$, i.e. $\theta = \pi/2$), in which case the classical ground state is macroscopically degenerate.

The model can be solved exactly employing the technique developed in a seminal paper by Lieb, Schultz and Mattis~\cite{LIEB1961407} and then applied to the transverse field Ising chain in ~\cite{Pfeuty:1970qrn}.
The first step is to map the spin operators onto fermionic creation and annihilation operators, $\hat{c}$ and $\hat{c}^{\dagger}$, by carrying out the Jordan-Wigner transformation~\cite{Jordan1928berDP} defined as
\be
\begin{split}
    \hat{\sigma}^z_j &= e^{i\pi\sum_{k = 1}^{j-1} \hat{c}_k^{\dagger}\hat{c}_k}\left(\hat{c}_j^{\dagger} + \hat{c}_j\right)\\
    \hat{\sigma}^y_j &= i e^{i\pi\sum_{k = 1}^{j-1} \hat{c}_k^{\dagger}\hat{c}_k}\left(\hat{c} - \hat{c}^{\dagger} \right)\\
    \hat{\sigma}^x_j &= 2\hat{c}_j^{\dagger}\hat{c} - 1.\\
\end{split} \label{JW_trafo}
\ee

Imposing open boundary conditions, we express the Hamiltonian [Eq.\eqref{tkc_hamilt_spins}] in terms of spinless fermions:
\be
\begin{split}
    \hat{H} = &-J\sum_{j = 1}^L \left(1 + (-1)^j\gamma\right)\left[ \left(\cos{2\theta} - i(-1)^j\sin{2\theta}\right)\hat{c}_j^{\dagger}\hat{c}_{j+1}^{\dagger}\right. \\& \left. - \left(\cos{2\theta} + i(-1)^j\sin{2\theta}\right)\hat{c}_j\hat{c}_{j+1}\right. + \left.\hat{c}_j^{\dagger}\hat{c}_{j+1} - \hat{c}_j \hat{c}_{j+1}^{\dagger} \right]\\
    &- h\sum_{j = 1}^{L} \left( \hat{c}_j^{\dagger}\hat{c} - \hat{c}_j \hat{c}^{\dagger} \right).
\end{split}
\ee

For later purposes, it is useful to introduce a column vector $\mathbf{\Psi}$ of length $2L$ 
\be
\mathbf{\Psi}^{\dagger} = (\hat{c}_1^{\dagger},~\hat{c}_2^{\dagger},..., \hat{c}_L^{\dagger},~\hat{c}_1,~\hat{c}_2,..,\hat{c}_L)
\ee
and rewrite the Hamiltonian in the form showing its particle-hole symmetry:
\be
\hat{H} = \mathbf{\Psi}^{\dagger}\begin{pmatrix}
    A & B\\
    -B^{*} & -A^{*}\\
\end{pmatrix}\mathbf{\Psi}
\ee
with
\be
\begin{split}
    A &= -\frac{J}{2}\left(1 + (-1)^j\gamma\right)\left(\delta_{j,j+1} +   \delta_{j+1,j}\right) - h\delta_{j,j}\\
    B &= -\frac{J}{2}\left(1 + (-1)^j\gamma\right)\left(\cos{2\theta} - i(-1)^j\sin{2\theta} \right)\\
    &\times\left(\delta_{j,j+1} - \delta_{j+1,j}  \right).
\end{split}
\ee

Next, we diagonalize the Hamiltonian using a Bogoliubov transformation~\cite{Bogoljubov1958OnAN}. One can notice that if $(u_{\mu},~v_{\mu})^T$ is an eigenvector with the eigenvalue $\varepsilon_{\mu}$, $(v_{\mu}^{*},~v_{\mu}^{*})^T$ is an eigenvector with the eigenvalue $-\varepsilon_{\mu}$. Hence, we can organize the eigenvectors into the following matrix
\be
\mathcal{U} = \begin{pmatrix}
    u_1 & ... & u_L & v_1^{*} &... & v_L^{*}\\
    v_1 & ... & v_L & u_1^{*} & ... & u_L^{*}\\
\end{pmatrix} = \begin{pmatrix}
    U & V^{*}\\
    V & U^{*}\\
\end{pmatrix}. \label{eigenvectors}
\ee
After introducing a Nambu vector $\mathbf{\Phi}$ as $\mathbf{\Psi} = \mathcal{U}\mathbf{\Phi}$, we can write the Hamiltonian in a diagonal form 
\be
\hat{H} =  \mathbf{\Phi}^{\dagger}\hat{E} \mathbf{\Phi}~\text{with}~E = \text{diag}(\ve_1,..., \ve_L,~-\ve_1,..., -\ve_L).
\ee
The Bogoliubov operators $\hat{\eta}^{\dagger}$ and $\hat{\eta}$ are related to the initial fermion operators as follows:
\begin{equation}
    \hat{c}_i = U_{ij}\hat{\eta}_j + V^{*}_{ij}\hat{\eta}^{\dagger}_j,~~
    \hat{c}^{\dagger}_i = V_{ij} \hat{\eta}_j + \hat{U}^{*}_{ij}\hat{\eta}^{\dagger}_j.\label{Bogoliubov}
\end{equation}

Although the energy spectrum can be easily obtained numerically, it is also instructive to do it analytically assuming periodic boundary conditions and going into momentum representation. Note that in the case of periodic b.c. there are no zero-energy edge excitations, which are present in the ferromagnetic phase in the case of the open boundary conditions. Additionally, this approach cannot be extended to the disordered case, which will be examined later in this paper. For this reason, we will be using the open b.c. in the following.

Leaving out technical details, which are provided in Appendix \ref{eps_tkc}, here we outline the most important result for the case with periodic b.c., namely, that the quantum phase transition (QPT) occurs at the magnetic field strength given by
\be
h_c/J = \sqrt{1 - \gamma^2}|\cos{2\theta}|.
\ee
At $h < h_c$, the spins form a ferromagnet, which is polarized in the $z$-direction, and for $h > h_c$ the system is in a paramagnetic phase.

\section{Calculation of the dynamical structure factor}
The dynamical structure factor can be obtained from the time-dependent spin-spin correlation functions as follows:
\be
S^{\alpha\beta}(\omega, k) = \frac{1}{L}\int_{-\infty}^{+\infty}dt\,\sum_{i,j}e^{ik(i - j)}e^{i\omega t}\,\langle \hat{\sigma}_i^\alpha(t)\hat{\sigma}_j^{\beta}(0)\rangle, \label{S_def_general}
\ee
where $\hat{\sigma}_i^\alpha(t) = e^{i\hat{H}t}\hat{\sigma}_i^\alpha(0)e^{-i\hat{H}t}$, $L$ is the total number of sites, and $\langle ... \rangle$ is the expectation value in the ground state.

For the case with translation symmetry, the expression above can be simplified so that we could avoid calculating $L^2$ correlators. Specifically, we use that $\langle \hat{\sigma}_i^\alpha(t)\hat{\sigma}_j^{\beta}(0)\rangle = \langle \hat{\sigma}_{i + 2n}^\alpha(t)\hat{\sigma}_{j+ 2n}^{\beta}(0)\rangle$ for $n = 0,1,2...$ and rewrite the expression above as follows
\be
S^{\alpha\beta}(\omega, k) = \int_{-\infty}^{+\infty}dt\,\sum_r e^{ikr}e^{i\omega t}\,C^{\alpha,\beta}(r,t) 
\ee
with
\be
\begin{split}
C^{\alpha,\beta}(r,t) &= \frac{1}{2}\left(\langle \hat{\sigma}_{2i}^\alpha(t)\hat{\sigma}_{2i+r}^{\beta}(0)\rangle \right.\\
&+\left. \langle \hat{\sigma}_{2i + 1}^\alpha(t)\hat{\sigma}_{2i + 1 + r}^{\beta}(0)\rangle\right).\\
\end{split}
\ee
 
In our numerical calculations we work with finite systems which exhibit finite size effects, so in order to reduce the influence of these, we choose $i$ such that the corresponding sites ($2i$ and $2i + r$, or $2i + 1$ and $2i + r + 1$) are close to the middle of the chain.\\

\subsection{Correlation functions in terms of Pfaffians}

The next step is to calculate the above mentioned correlation functions.
Following \cite{PhysRevB.108.155143} and \cite{PhysRevB.107.045124}, we employ the fermionic representation and express correlators in terms of Pfaffians~\cite{Cayley1849}.

We further focus on the calculation of $S^{zz}(\omega, k)$, although other spin components can be obtained in a similar manner. Applying the Jordan-Wigner transformation (Eq.\eqref{JW_trafo}) and using the fact that 
\be
e^{i\pi c_i^{\dagger}c_i} = (c_i^{\dagger} + c_i)(c_i^{\dagger} - c_i),
\ee
we rewrite the corresponding correlation functions as follows
\be
\begin{split}
\langle \hat{\sigma}_i^{z}(t) \hat{\sigma}_j^{z}(0) \rangle =& \langle \prod_{k = 1}^{i-1} (\hat{A}_k(t)\hat{B}_k(t))\hat{A}_i(t)\\
\times&\prod_{m = 1}^{j-1}(\hat{A}_m(0)\hat{B}_m(0))\hat{A}_j(0)\rangle,\\
\end{split}\label{corr_ferm}
\ee 
with
\be
\begin{split}
    \hat{A}_j(t) &= \hat{c}_j^{\dagger}(t) + \hat{c}_j(t),\\
    \hat{B}_j(t) &= \hat{c}_j^{\dagger}(t) - \hat{c}_j(t).\\
\end{split} \label{AB_def}
\ee

Using Wick's theorem, we can rewrite Eq.\eqref{corr_ferm} as a Pfaffian of a $2(i + j -1)$-dimensional skew-symmetric matrix~\cite{mccoy2014two}:
\be
\langle \hat{\sigma}_i^{z}(t) \hat{\sigma}_j^{z}(0) \rangle = \Pf S
\ee
with
\begin{widetext}
\be
S = \begin{pmatrix}
    0 & \langle A_1(t)B_1(t)\rangle & \langle A_1(t)A_2(t)\rangle & \langle A_1(t)A_2(t)\rangle & ... & \langle A_1(t)A_j(0)\rangle\\
    -\langle A_1(t)B_1(t)\rangle & 0 & \langle B_1(t)A_2(t)\rangle &
    \langle B_1(t)B_2(t)\rangle & ... & \langle B_1(t)A_j(0)\rangle\\
    -\langle A_1(t)A_2(t) & -\langle B_1(t)A_2(t)\rangle  & 0 & \langle A_2(t)B_1(t)\rangle & ... & \langle A_2(t)A_j(0)\rangle \\
    \vdots & \vdots & \vdots & \vdots & \ddots & \vdots\\
    -\langle A_1(t)A_j(0)\rangle & -\langle B_1(t)A_j(0)\rangle & -\langle A_2(t)A_j(0)\rangle & -\langle B_2(t)A_j(0)\rangle &...&0\\
\end{pmatrix}. \label{pfaff_matrix}
\ee
\end{widetext}

To calculate the entries of the matrix, we carry out the Bogoliubov transformation defined in Eq.\eqref{Bogoliubov} and use the identity $\langle \hat{\eta}_m(t)\hat{\eta}_n^{\dagger}(0) \rangle = e^{-i\varepsilon_m t}\delta_{mn}$. After some algebra, we obtain the expressions for the entries
\be
\begin{split}
    \langle \hat{A}_i(t)\hat{A}_j(0)\rangle &= \sum_{m}e^{-i\eps_m t}(U_{im} + V_{im})(U_{jm}^{*} + V_{jm}^{*}),\\
    \langle \hat{A}_i(t)\hat{B}_j(0)\rangle &= \sum_{m}e^{-i\eps_m t}(U_{im} + V_{im})(U_{jm}^{*} - V_{jm}^{*}),\\
    \langle \hat{B}_i(t)\hat{A}_j(0)\rangle &= \sum_{m}e^{-i\eps_m t}(V_{im} - U_{im})(U_{jm}^{*} + V_{jm}^{*}),\\
    \langle \hat{B}_i(t)\hat{B}_j(0)\rangle &= \sum_{m}e^{-i\eps_m t}(V_{im} - U_{im})(U_{jm}^{*} - V_{jm}^{*}).\\
\end{split}
\ee
The unitary matrices, $\hat{U}$ and $\hat{V}$, and the eigenvalues $\varepsilon_m$ can be calculated numerically.

\subsection{Numerical evaluation of the structure factor} \label{numerics}
In this subsection, we elaborate on the technical details for the calculation of the correlation functions and structure factor. We follow previous work on the 1D TFIM by Derzhko and Khrokhmalskii~\cite{PhysRevB.56.11659} and cover here points  their implementation.

The efficient computation of Pfaffians has been a long-standing challenge. Typically, it can be addressed by making use of the identity $(\Pf{A})^2 = \det A$. However, determining the sign after taking the square root is the inherent complication~\cite{knolle2015dynamics}. Fortunately, recent developments in the numerical evaluation of Pfaffians~\cite{Wimmer_2012} have improved efficiency and, therefore, we use the Python library "Pfapack"~\cite{pfapack} in our calculations.

Another computational issue arising in the ferromagnetic phase is related to the presence of zero-energy eigenvalues~\cite{Mbeng_2024}. Since the zero-energy subspace is two-fold degenerate, the eigenvectors obtained by an exact diagonalization routine do not necessarily have the form shown in Eq.\eqref{eigenvectors}. Thus, we need to enforce this structure explicitly. The corresponding technical details can be found in Appendix~\ref{zero_modes}.

Next, we work with finite systems, which inherently exhibit finite-size effects. Therefore, we choose the largest system size that still allows us to run simulations in a reasonable time, e.g. systems with a few hundred sites turned out to be a good compromise. In addition, the presence of finite-size effects should be taken into account when processing the data, for example, in the numerical routine for the Fourier transform. See Appendix~\ref{finite_size} for more details.

Finally, at $h < h_c$, the ferromagnetic order leads to the emergence of a delta function peak at $S(k,\omega = 0)$, with a dominating intensity such that other features become barely discernible. We eliminate the delta peak by subtracting the ground state magnetization. The details can be found in Appendix~\ref{delta_peak}.

\section{Dynamic structure factor of the twisted Kitaev chain}

\begin{figure*}
    \centering
     \begin{subfigure}[b]{0.28\paperwidth}
    \centering
        \includegraphics[width=6.4 cm]{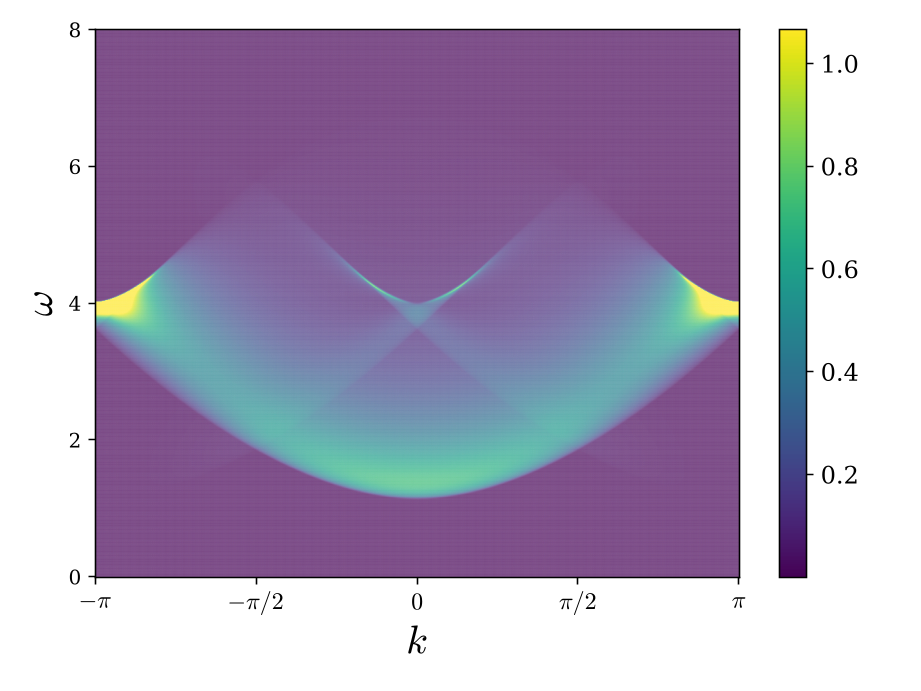} 
        \caption{} \label{Fig2:tkc_ferro}
    \end{subfigure}
    \begin{subfigure}[b]{0.28\paperwidth}
          \centering
        \includegraphics[width=6.4  cm]{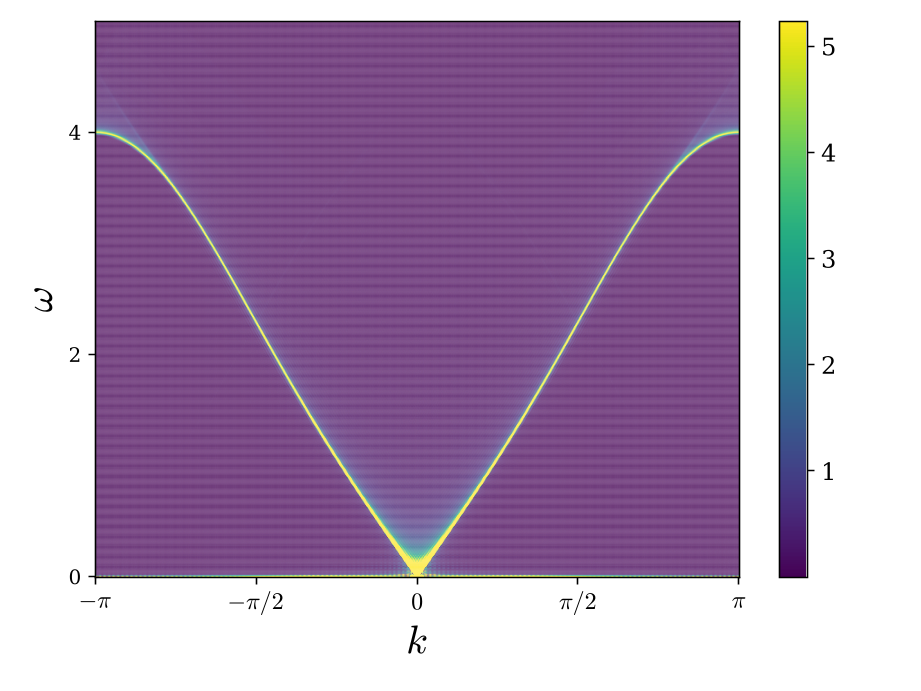} 
        \caption{} \label{Fig2:tkc_crit}
    \end{subfigure}
    \begin{subfigure}[b]{0.28\paperwidth}
       \centering
        \includegraphics[width= 6.4 cm]{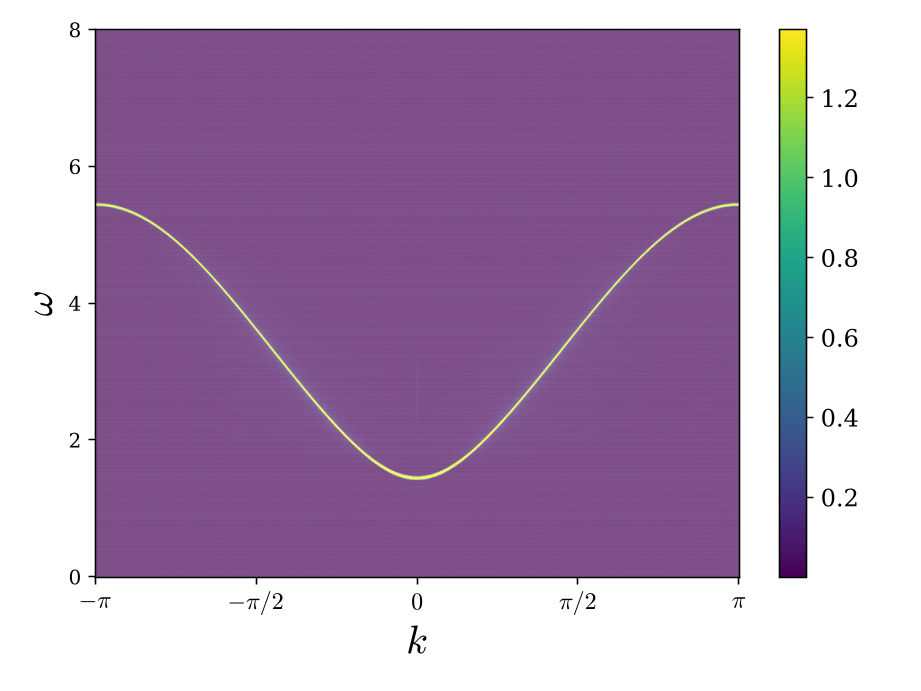} 
        \caption{} \label{Fig2:tkc_param}
    \end{subfigure}
    \caption{Structure factor $S(k,\omega)$ for the TKSC in the ferromagnetic phase, $h = 0.5h_c$ (\subref{Fig2:tkc_ferro}), at the critical point, $h = h_c$ (\subref{Fig2:tkc_crit}), in the paramagnetic phase, $h = 2h_c$ (\subref{Fig2:tkc_param}). For all of the figures we chose $\theta = \pi/10$ and $L = 500$.}\label{tkc_3_phases}
\end{figure*}
First, we focus on the conventional TKSC, namely without glide symmetry breaking, which corresponds to the case of $\gamma = 0$. We choose $\theta = \pi/10$ to be close to the angle observed in experiments with \ch{Co Nb_2 O_6}~\cite{Morris_2021}, and examine both the ferromagnetic and paramagnetic phases, as well as the quantum critical point (QCP).

The results for the structure factor in the ferromagnetic phase are shown in Fig.\ref{tkc_3_phases}\subref{Fig2:tkc_ferro}. In order to understand the origin of the quasiparticle continuum we recall that in the ferromagnetic phase the excitations above the ground state can be interpreted as domain walls~\cite{PhysRevLett.78.2220}. However, spin flips probed via the dynamical structure factor excite pairs of domain walls and, consequently, one does not observe single-particle excitations, but a quasiparticle continuum. The main contribution to the latter arises from two-quasiparticle excitations, as in the classic TFIM.

\begin{figure}
    \centering
        \includegraphics[width= 8.0 cm]{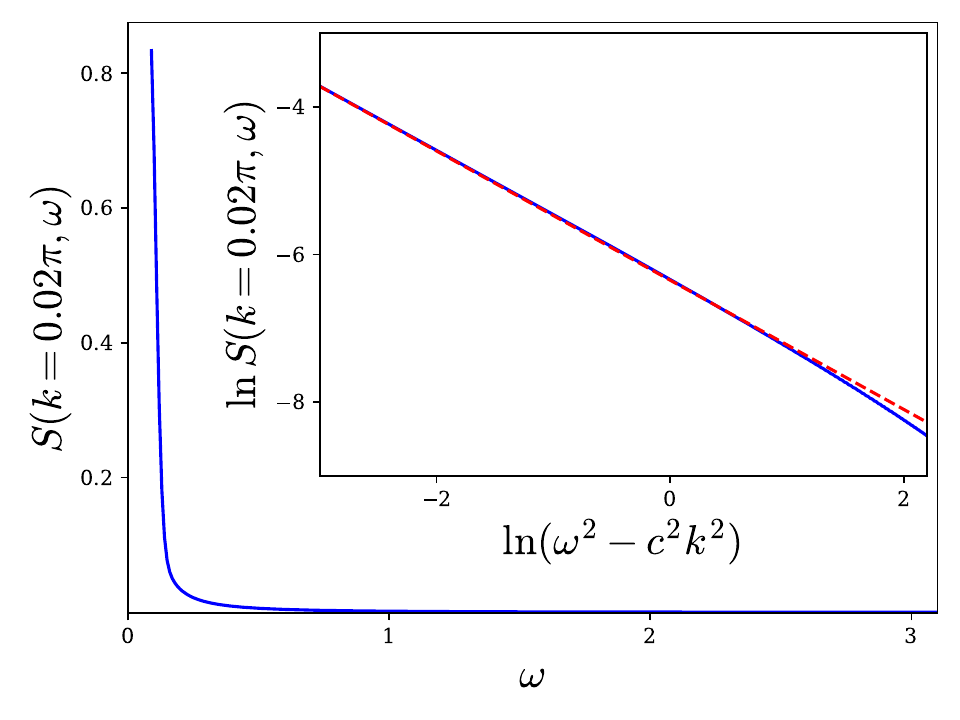} 
    \caption{ Frequency dependence of the structure factor $S(k,\omega)$ at $k = 0.02\pi$ for the TKSC at the critical point. The inset shows a log-log plot at $k = 0.02\pi$. The fit $y = -0.875x - 6.35$ (red dashed line) is close to the anomalous exponent $\eta = 1/4$. The figures correspond to $\theta = \pi/10$, $h = h_c$, $L = 500$.}\label{tkc_crit_point}
\end{figure}

The structure factor at the quantum critical point is shown in Fig.\ref{tkc_3_phases}\subref{Fig2:tkc_crit}. There are no well-defined quasiparticle excitations, but the continuum of critical excitations~\cite{sachdev2011}. Indeed, even though one can observe the bright lines corresponding to the quasiparticle excitation spectrum, these lines are, in fact, blurred, and rather the onset of a whole continuum of excitations.
This quantum critical behaviour can be examined quantitatively by looking at a fixed momentum (see Fig.\ref{tkc_crit_point}). The frequency dependence of the structure factor can be compared with the field theory calculations~\cite{sachdev2011}, which yield the following asymptotic predictions for small $k$ and low $\omega$:
\be
S(k,\omega) \sim \frac{\sign(\omega) \theta(|\omega| - ck)}{(\omega^2 - c^2k^2)^{1-\eta/2}}\label{S_eta}
\ee
with $c$ given by Eq.\eqref{low_energ_spectrum} and the anomalous dimension $\eta$. For the TFIM $\eta = 1/4$~\cite{sachdev2011} and since our TKSC falls in the same universality class we expect a similar behavior. The inset in Fig.\ref{tkc_crit_point} shows the corresponding log-log plot, from which we can extract the value of $\eta$. We find from fitting the numerical curve, the value close to $\eta = 1/4$ which confirms that the TKSC belongs to the Ising universality class. Note, we chose a small but non-zero value of $k$ to reduce the influence of finite-size effects (see Appendix \ref{finite_size}).

In Fig.\ref{tkc_3_phases}\subref{Fig2:tkc_param} we present the results for the paramagnetic phase. Here, the structure factor is dominated by single-quasiparticle excitations. Indeed, for $h \gg h_c$, all spins in the ground state are aligned in the magnetic field direction, and quasiparticle excitations correspond to single-spin flips. The absence of broadening reflects the stability of these excitations as expected from the strong-coupling and field theory calculations~\cite{sachdev2011}. The latter also predicts the existence of a three-particle continuum, which we do have, but whose contribution is too small to be observed on top of the single-particle mode. Alongside the main mode, the excitation spectrum also includes a similar mode shifted by $\delta k = \pi$, originating from the chain buckling that results in the unit cell doubling. However, since the buckling at $\theta = \pi/10$ is relatively small, this ``shadow'' mode has very low intensity ~\cite{PhysRevB.90.014418} and is not visible in the figure for the structure factor.

\begin{figure}
    \centering
        \includegraphics[width=9.5 cm]{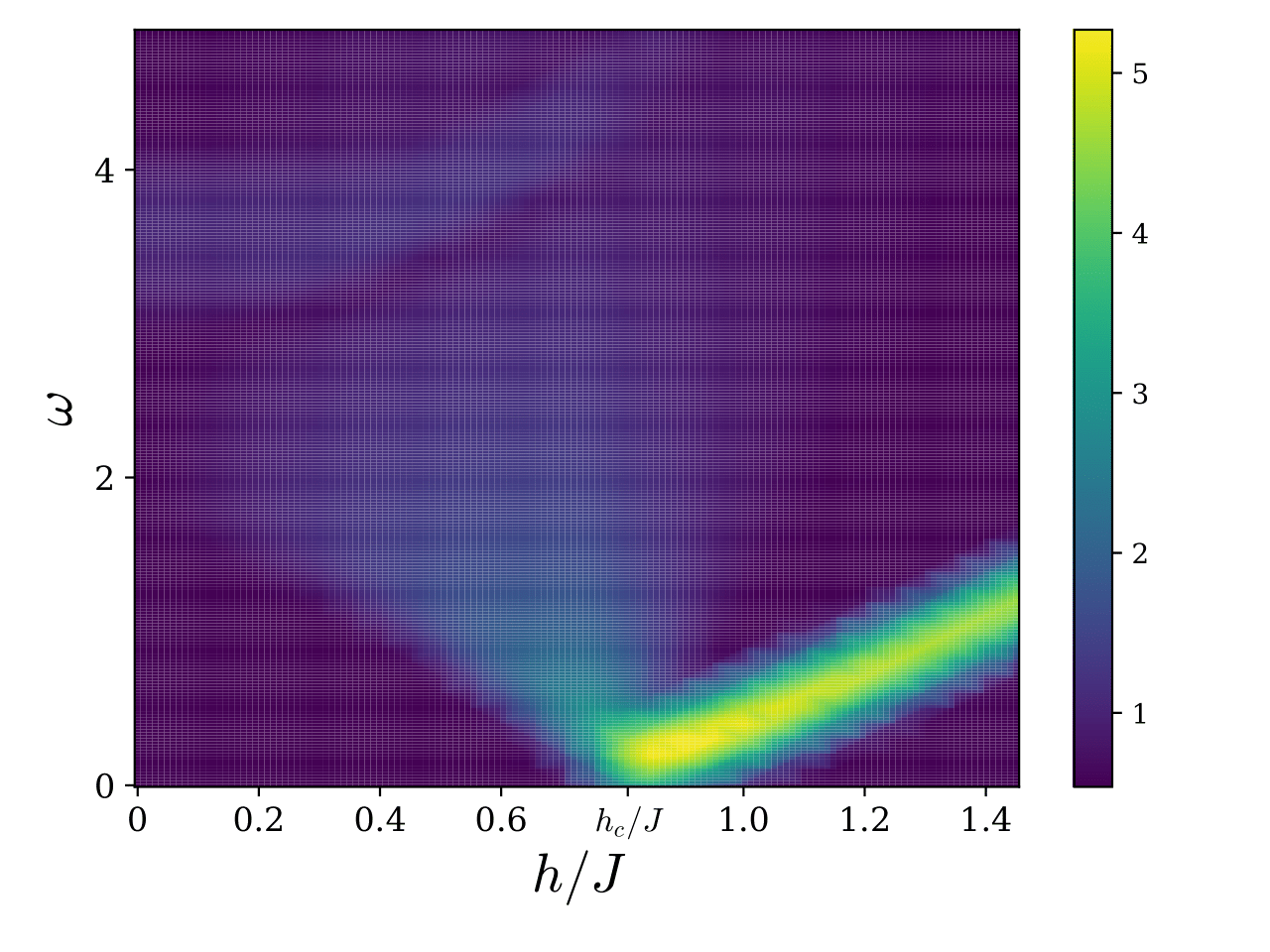} 
    \caption{The evolution of $S(k = 0,\omega)$ for the TKSC across the quantum critical point. $\theta = \pi/10$, $L = 100$, $h_c/J = \cos{2\theta}$.}\label{tkc_field}
\end{figure}

To make a connection with experiments which tune magnetic field across the critical point (see, for example,~\cite{Morris_2021}), we show the evolution of the dynamic structure factor at $k = 0$ across the critical point (see Fig.\ref{tkc_field}). One can notice that the calculated structure factor is qualitatively similar to the experimental results obtained by time-domain THz spectroscopy~\cite{Morris_2021}. The finite size effects -- our calculations were performed for a system of 100 sites -- account for the fact that paramagnetic peak appears to be broader than in Fig.\ref{tkc_3_phases}\subref{Fig2:tkc_param}.

\section{TKSC with broken glide symmetry}
\begin{figure*}
    \centering
     \begin{subfigure}[b]{0.28\paperwidth}
    \centering
        \includegraphics[width=6.4 cm]{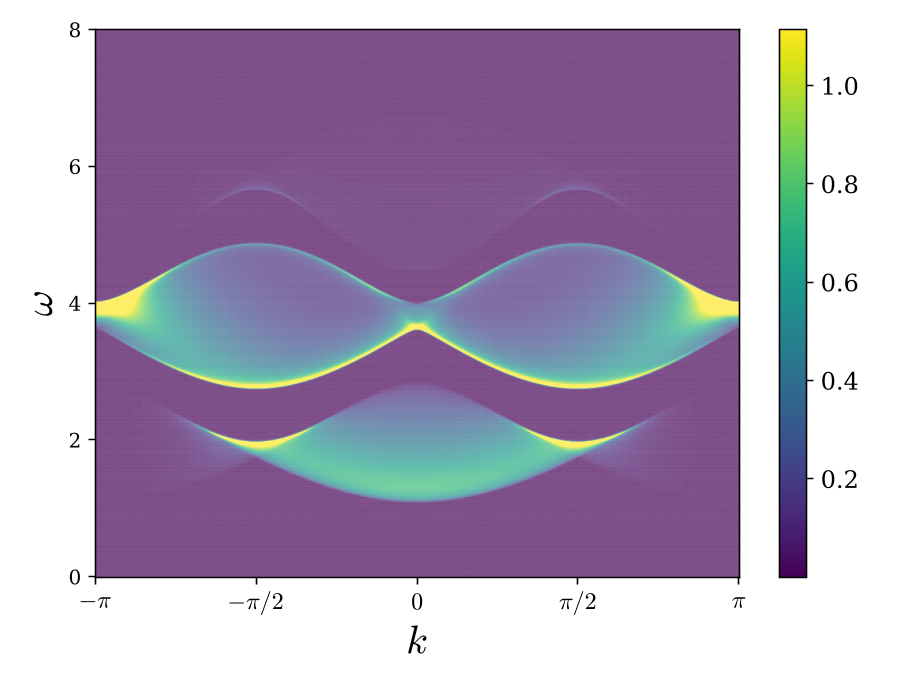} 
        \caption{} \label{tkc_ferro_broken}
    \end{subfigure}
    \begin{subfigure}[b]{0.28\paperwidth}
          \centering
        \includegraphics[width=6.4  cm]{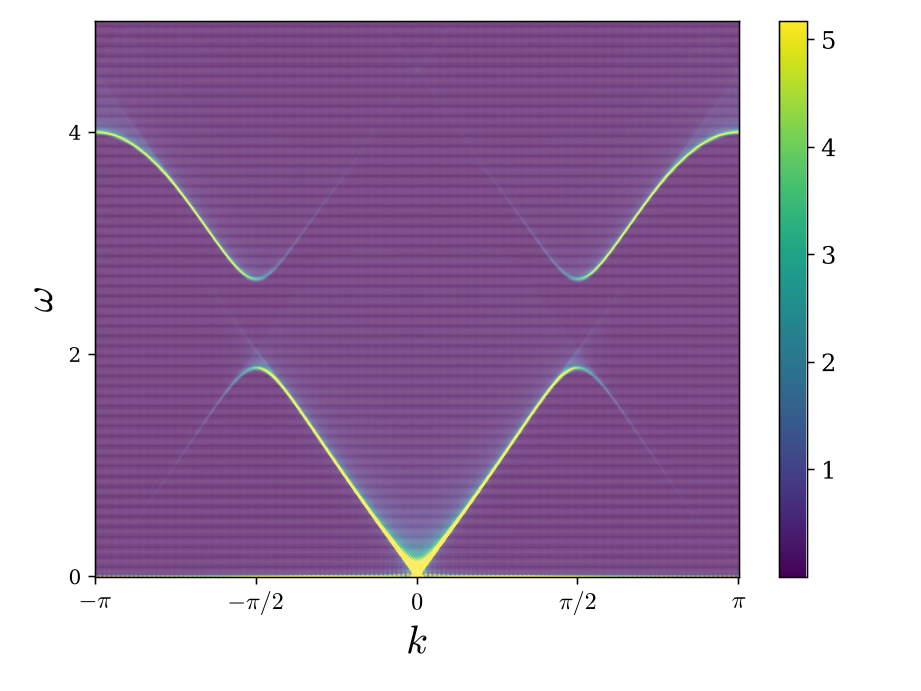} 
        \caption{} \label{tkc_crit_broken}
    \end{subfigure}
    \begin{subfigure}[b]{0.28\paperwidth}
       \centering
        \includegraphics[width= 6.4 cm]{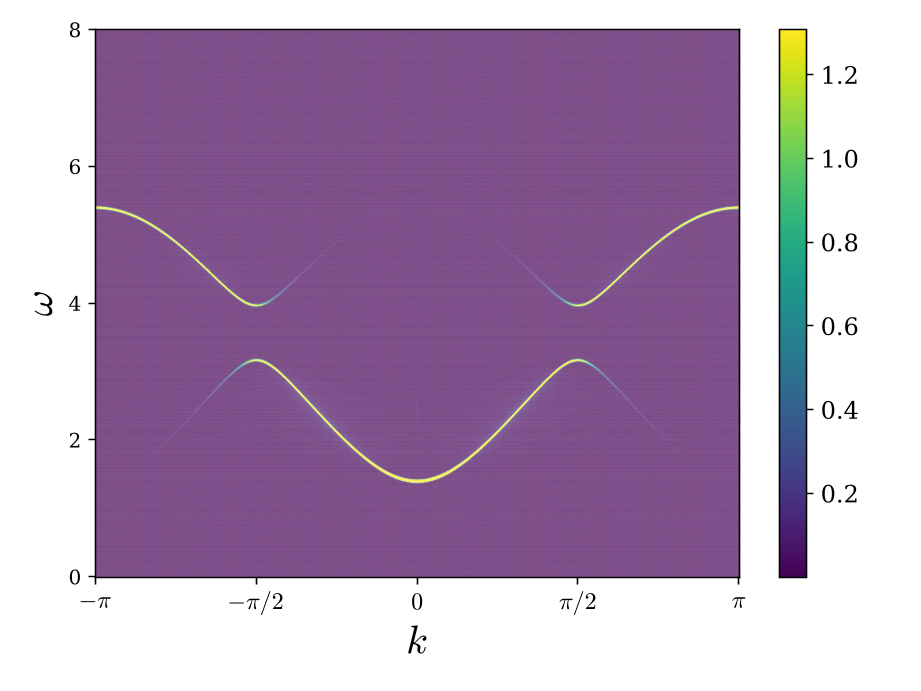} 
        \caption{} \label{tkc_param_broken}
    \end{subfigure}
    \caption{ The structure factor $S(k,\omega)$ for the TKSC with broken glide symmetry in the ferromagnetic phase, $h = 0.5h_c$ (\subref{tkc_ferro_broken}), at the critical point, $h = h_c$, (\subref{tkc_crit_broken}, in the paramagnetic phase, $h = 2h_c$ (\subref{tkc_param_broken}). The figures correspond to $\theta = \pi/10$, $\gamma = 0.2$, $L = 500$.}\label{tkc_broken_3_phases}
\end{figure*}

Here we examine the TKSC with broken glide symmetry, characterized by a non-zero parameter $\gamma$ in the Hamiltonian \eqref{tkc_hamilt_abstr_broken}. The glide symmetry breaking is evident in the emergence of a gap in the excitation spectrum at $k = \pm \pi/2$, see Fig.\ref{tkc_scheme} panel b). The structure factor is presented in Fig.\ref{tkc_broken_3_phases}, where one can see that the gap opening has a drastic effect on the high energy features. The low energy response can be understood using the same arguments presented for the TKSC without glide symmetry breaking. Note, that the gapping of the quasi-particle bands here leads to a much richer continuum response in the ferromagnetic phase. The gap observed at higher frequencies is directly set by the strength of the dimerization, e.g. quantifies the glide symmetry breaking.

\section{TKSC in presence of disordered magnetic field}
The same method allows us to study not only translational-invariant systems but also disordered ones, whose critical behavior remains  whose critical behavior is more subtle and continues to be actively investigated, particularly in low-dimensional settings~\cite{Rieger_1999}. In this section, we focus on the TKSC in a random magnetic field
\be
\begin{split}
\hat{H} = &-J\sum_{i = 1}^{L'}\left[\left( 1 + \gamma \right)\hat{\sigma}^{\hat{n}_1}_{2i - 1}\hat{\sigma}^{\hat{n}_1}_{2i} + \left( 1 - \gamma \right)\hat{\sigma}^{\hat{n}_2}_{2i}\hat{\sigma}^{\hat{n}_2}_{2i+1}\right]\\
&- \sum_{i = 1}^L h_i\hat{\sigma}^x_i, \label{tkc_hamilt_disordered}
\end{split}
\ee
where $h_i$ is drawn from a normal distribution
\be
\rho(h) = \frac{1}{\sqrt{2\pi\sigma^2}}e^{-\frac{(h - h_c)^2}{2\sigma^2h_c^2}}  \label{distribution}
\ee
with the mean $h_c$, which is taken to be the critical field for the system in a uniform magnetic field, and the variance $\sigma_h = \sigma h_c$.\\ 

To study the properties of systems with broken translational symmetry -- such as those in a disordered magnetic field (discussed in this section) and those in an incommensurate magnetic field (addressed in a later section) -- we continue to use the structure factor $S(k,\omega)$ defined in Eq.\eqref{S_def_general}. However, since $k$ is no longer a conserved quantum number, it should be understood as a parameter in the Fourier transform. We also note that, for a sufficiently large number of disorder realizations, translational symmetry is effectively restored on average~\cite{Jia_2006}, making $k$ a meaningful observable again.\\

\begin{figure}
    \centering
        \includegraphics[width=8 cm]{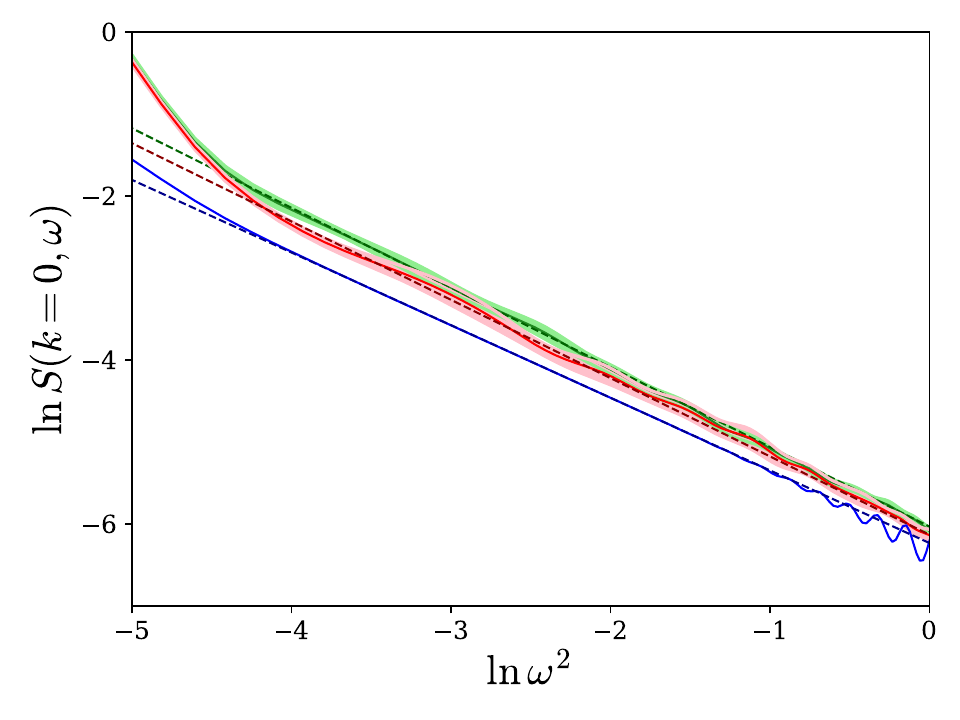} 
    \caption{Log-log plot of the frequency dependence of the structure factor $S(k,\omega)$ at $k = 0$ for the TKSC in a random transverse field (see Eq.\eqref{tkc_hamilt_disordered}). The blue solid line represents the results for the system without disorder; the blue dashed line indicates the corresponding linear fit ($y = -0.87x - 6.2$); the red and green solid curves show the average over 10 disorder realizations for the disordered models with $\sigma = 0.2$ and $\sigma = 0.1$, respectively; the red and green dashed lines indicate the corresponding linear fits ($y = -0.97x - 6.03$ for $\sigma = 0.2$ and $y = -0.96x - 6.13$ for $\sigma = 0.1$). The red and green shadings represent the standard errors of the average for each point on the red and green solid curves, respectively. $\overline{h} = h_c$, $\theta = \pi/6$, and $L = 200$.}\label{disord_exp_drift}
\end{figure}

Strictly speaking, it is not obvious whether the system described above indeed has a critical point at $\overline{h} = h_c = J\cos{2\theta}$. However, we  verified that this is the case for relatively small $\sigma \lesssim 0.2$ by measuring the ground state magnetization $\langle \hat{\sigma}^z\rangle$ and tuning the mean value of the magnetic field distribution across $h_c$.

Our goal here is to examine the quantum critical behavior of the system at the disordered quantum critical point. One might naively expect that for small variances $\sigma$, the structure factor at the critical point would still adhere to the scaling form  Eq.\eqref{S_eta}, perhaps with a different anomalous exponent. For example, in the case of spatial correlations the power-law decay of the clean system is modified to a new power-law exponent satisfying the Harris criterion~\cite{vojta2019disorder}. However, this is not the case for the dynamical behavior because  disorder breaks Lorentz symmetry -- i.e., the dynamical exponent $z$ is no longer equal to $1$~\cite{Young_1996}. Consequently, one cannot employ basic quantum to classical mapping and substitute $k^2 \rightarrow \omega^2 - c^2k^2$ in the classical result $S(k) \sim 1/k^{2-\eta}$ to obtain Eq.\eqref{S_eta}~\cite{sachdev2011}. Nevertheless, in Fig. \ref{disord_exp_drift}, we plot the disorder averaged low frequency part of the $k=0$ structure factor (red and green lines). Similar to the clean case (blue line), we find that an  analogous power-law behavior still persists for sizeable disorder characterized by $\sigma \lesssim 1$ (see Eq.\eqref{distribution}). While the numerical effort for obtaining these plots is considerable, we have checked that -- for other sufficiently small values of $\sigma$ such a power law divergence persists -- the exponent has no clear drift but rather stays at $\eta \approx 0$.

\section{TKSC in an incommensurate magnetic field}
Next, we consider the TKSC in a quasiperiodically modulated magnetic field, which can be construed as a highly-correlated random field: 
\be
\begin{split}
\hat{H} = &-J\sum_{j = 1}^{L'}\left[\left( 1 + \gamma \right)\hat{\sigma}^{\hat{n}_1}_{2j - 1}\hat{\sigma}^{\hat{n}_1}_{2j} + \left( 1 - \gamma \right)\hat{\sigma}^{\hat{n}_2}_{2j}\hat{\sigma}^{\hat{n}_2}_{2j+1}\right]\\
&- \sum_{j = 1}^L (h + A_h\cos(2\pi\alpha j))\hat{\sigma}^x_j, \label{tkc_hamilt_quasi}
\end{split}
\ee
with $\alpha = 2/(\sqrt{5}-1)$ and $h = J\cos(2\alpha)$. Note, other choices of irrational number $\alpha$ also work.
\begin{figure}
    \centering
        \includegraphics[width=8 cm]{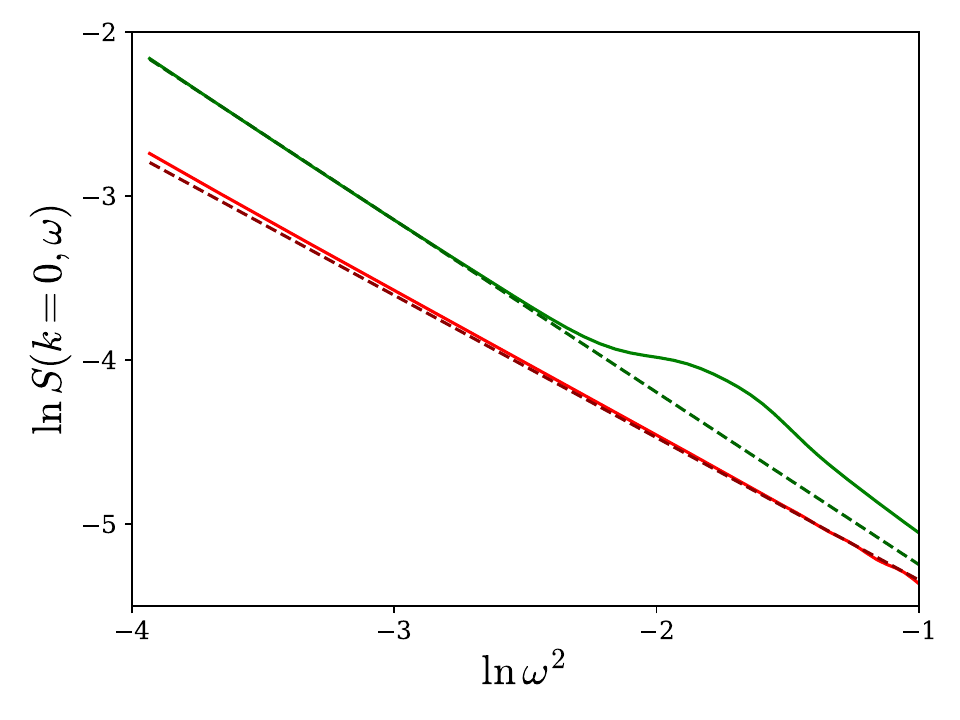} 
    \caption{ Log-log plot of the frequency dependence of the structure factor $S(k,\omega)$ at $k = 0$ for the TKSC in a incommensurate transverse magnetic field (see Eq.\eqref{tkc_hamilt_quasi}). The green solid line shows the results for $A_h = 0.2$, while the green dashed line indicates the corresponding fit ($y = - 1.0 x - 6.30$). The red solid curve represents the data for the pure system, i.e., $A_h = 0.0$, while the red dashed line indicated the corresponding fit ($y = - 0.87 x - 6.20$). $\overline{h} = h_c$, $\theta = \pi/6$, and $L = 200$.}\label{quasi_exp_drift}
\end{figure}

Similar to the model with a random field, the structure factor of the system at the critical point cannot be described by the scaling form Eq.\ref{S_eta}, but yet again we do observe a power-low behavior for relatively weak incommensurate fields ($A_h \lesssim 0.3$), see Fig.\ref{quasi_exp_drift}. Again we find $\eta\approx 0$.

In contrast to the model with a random field, one-dimensional quasiperiodic systems can demonstrate localization-delocalization phase transition~\cite{aubry_andre}. When the quasiperiodic modulation is weak, the system is in the extended phase with all eigenstates being delocalized. In the opposite case of strong modulation, the localized phase is observed. For intermediate localization strengths, the system enters the intermediate phase characterized by the coexistence of localized and delocalized states separated by mobility edges~\cite{PhysRevB.96.085119}.

\begin{figure}
    \raggedright
        \includegraphics[width=8 cm]{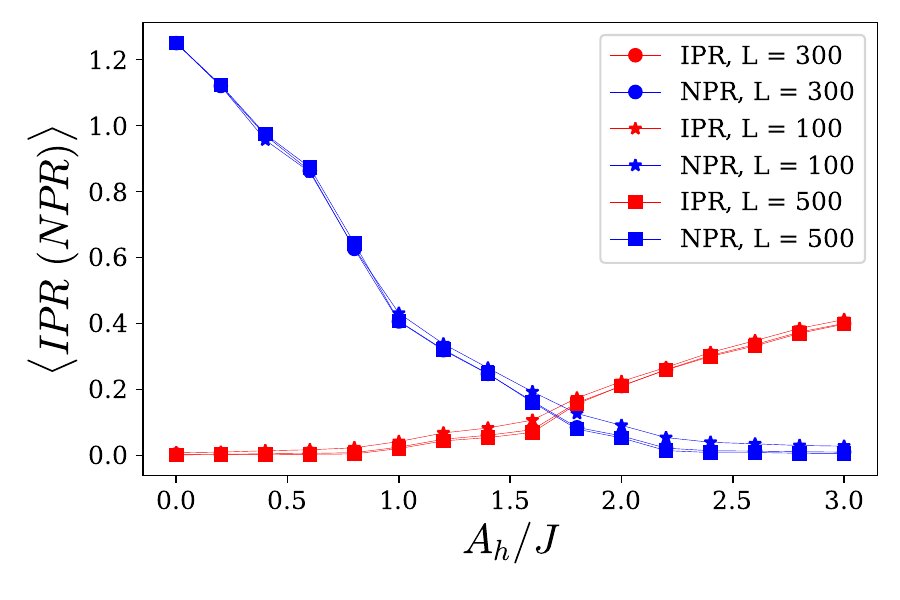} 
    \caption{IPR (NPR) for the TKSC in an incommensurate magnetic field (see Eq.\eqref{tkc_hamilt_quasi}) as a function of the modulation strength $A_h$ for different system sizes: $L = 100$, $L = 300$, and $L = 500$. }\label{tkc_quasi_ipr}
\end{figure}

A specific localization phase can be diagnosed by the simultaneous calculation of the average inverse participation ratio (IPR) and the average normalized participation ratio (NPR) over all single-particle eigenstates ~\cite{PhysRevB.96.085119},~\cite{PhysRevB.101.064203}. For the $i$-th eigenstate $u_n^{(i)}$, they are defined as 
\be
\text{IPR}^{(i)} = \sum_n |u_n^{(i)}|^4,~~\text{NPR}^{(i)} = \left[L\sum_n |u_n^{(i)}|^4\right]^{-1}.
\ee
In the extended phase, the IPR is finite while the NPR is vanishing as ($1/L$). Conversely, in the localized phase, the NPR is finite and IPR is vanishing. The intermediate phase is characterized by finite values of both IPR and NPR.\\

The results of the corresponding calculation for the system with the Hamiltonian \eqref{tkc_hamilt_quasi} are shown in Fig.\ref{tkc_quasi_ipr}. These results indicate that the extended phase is observed for $A_h/J \lesssim 1.0$, the intermediate phase for  $A_h/J \sim 1.0-2.0$, and the localized phase for $A_h/J \sim 2.0$. However, it is evident that the transition values determined via this method have significant error bars. Consequently, for further studies we select modulation values far from the phase transition boundaries.

\begin{figure*}
    \centering
     \begin{subfigure}[b]{0.28\paperwidth}
    \centering
        \includegraphics[width=6.4 cm]{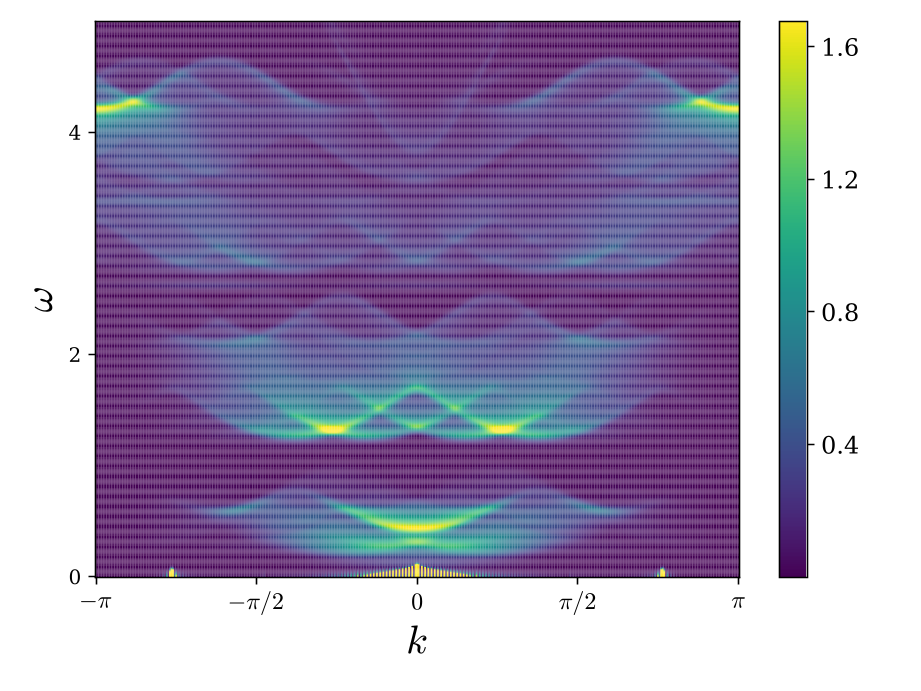} 
        \caption{} \label{tkc_quasi_0.5}
    \end{subfigure}
    \begin{subfigure}[b]{0.28\paperwidth}
          \centering
        \includegraphics[width=6.4  cm]{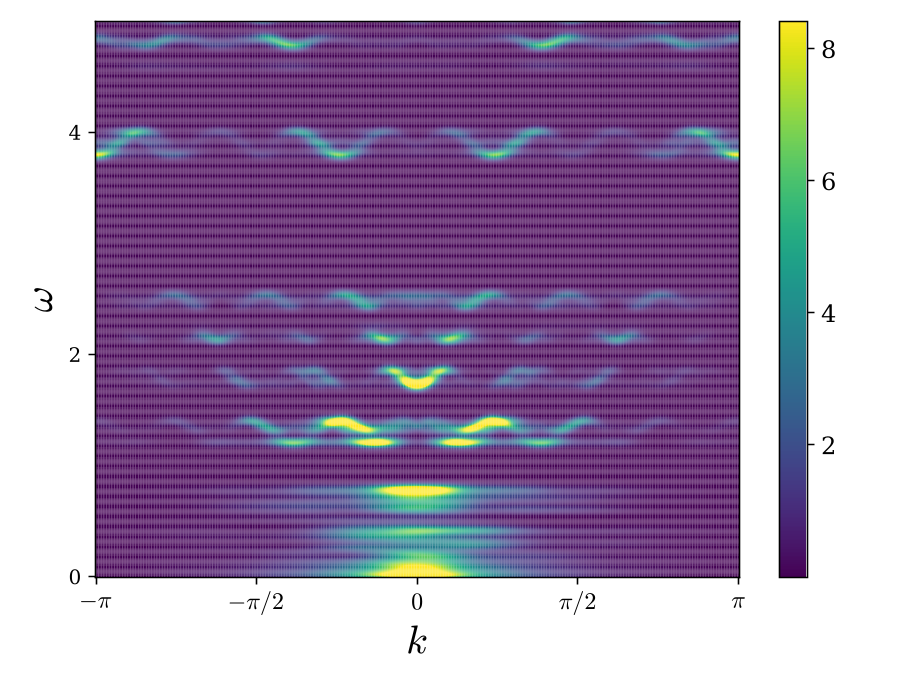} 
        \caption{} \label{tkc_quasi_1.5}
    \end{subfigure}
    \begin{subfigure}[b]{0.28\paperwidth}
       \centering
        \includegraphics[width= 6.4 cm]{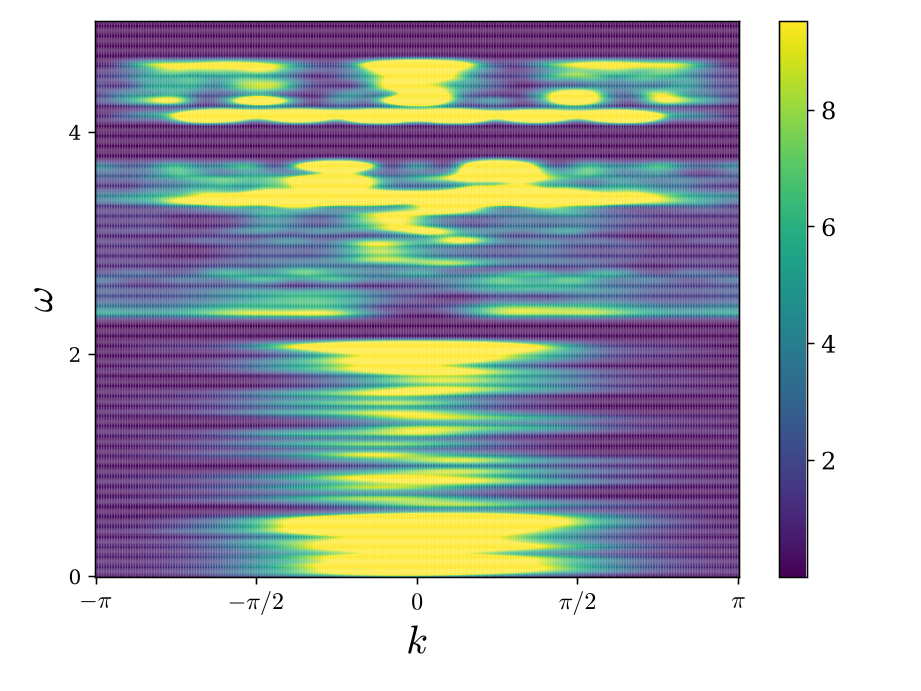} 
        \caption{} \label{tkc_quasi_3.0}
    \end{subfigure}
    \caption{The structure factor for the TKSC in an incommensurate magnetic field (see Eq.\eqref{tkc_hamilt_quasi}) in the \subref{tkc_quasi_0.5}) delocalized phase ($A_h = 0.5$) \subref{tkc_quasi_1.5}) intermediate phase ($A_h = 1.5$) \subref{tkc_quasi_3.0}) localized phase ($A_h = 3.0$). The plots correspond to $\theta = \pi/6$, $\overline{h} = h_c$, $L = 200$.}\label{tkc_incom_loc_deloc}
\end{figure*}

In Fig.\ref{tkc_incom_loc_deloc} we present the structure factor for the TKSC in a quasiperiodic magnetic field across all three phases discussed above. In each phase, the correlated disorder induces gaps in the spectrum, leading to the formation of minibands in the two-quasiparticle response~\cite{PhysRevB.41.5544}.  In both the localized and intermediate phases, panels b) and c), one can observe the horizontal stripe-like patterns independent of momentum, which indicate the localized modes~\cite{PhysRevB.74.172414}. In constrast,  the wave-like patterns in the delocalized and intermediate phase at high frequency show a clear momentum dependence  signifying the presence of delocalized modes. Notably, the results for the intermediate phase reveal the presence of a single-particle mobility edge. Indeed, if all single-particle states below a certain threshold are localized, a similar threshold should appear in the two-quasiparticle response, which is exactly what is observed in Fig.\ref{tkc_incom_loc_deloc}\subref{tkc_quasi_1.5}.

\section{Discussion and Conclusions}
The TKSC provides a minimal soluble model for understanding the behavior Ising spin chain materials like  $\ch{CoNb2O6}$. Our analysis of the TKSC has yielded exact predictions for the dynamical structure factor, encompassing both the universal low-energy quantum critical features and the non-universal high-frequency phenomena. The results shown in Fig.\ref{tkc_field} are in good agreement with the experimental findings presented in Fig. 3b of the paper by Morris et al.~\cite{Morris_2021}. However, some high-frequency features observed in Fig. 3a of the same paper were not reproduced within the framework of the TKSC. This discrepancy suggests that the Hamiltonian governing the behaviour of $\ch{CoNb2O6}$ might include some additional terms beyond those considered in the TKSC or even to go beyond the independent chain approximation. Apart from the quantum materials realization, other platforms exist. For example, the random transverse field Ising model has been proposed to be experimentally realized in artificial spin ice~\cite{PhysRevLett.127.207203}, and quasiperiodic potentials naturally arise in cold atom experiments using lasers with incommensurate wave vectors~\cite{PhysRevLett.103.013901}. Thus, these systems could serve as promising starting point for the experimental realization of similar TKSC-like models.

From the theoretical perspective our results provide rigorous examples of dynamical quantum critical response functions beyond the seminal TFIM chain. In addition, beyond the clean systems we have presented first results for the TKSC in a random transverse field and in a quasiperiodic transverse field. Understanding from a field theory point of view the appearance of the low frequency power-law divergence induced by disorder and the change of critical exponents in frequency for a system without Lorentz symmetry remains an open challenge.


\acknowledgments
 U.K. would like to thank Markus Drescher for the productive discussions that greatly helped with the numerical calculations and Anubhav Srivastava for an insightful discussion on the glide symmetry breaking.

 J.K. thanks the hospitality of the Aspen Center for Physics,
which is supported by National Science Foundation
grant PHY-2210452. JK acknowledges support from the Deutsche Forschungsgemeinschaft (DFG, German Research Foundation) under Germany’s Excellence Strategy– EXC–2111–390814868 and DFG Grants No. KN1254/1-2, KN1254/2-1 and TRR 360 - 492547816, as well as the Munich Quantum Valley, which is supported by the Bavarian state government with funds from the Hightech Agenda Bayern Plus. J.K. further acknowledges support from the Imperial-TUM flagship partnership.

D.K. acknowledges support from Labex MME-DII grant ANR11-LBX-0023, and funding under The Paris Seine Initiative Emergence programme 2019. 

\bibliography{biblio_tkc}

\appendix
\section{Energy spectrum of the TKC} \label{eps_tkc}
We start from the Hamiltonian [Eq.\eqref{tkc_hamilt_spins}], but this time we impose periodic boundary conditions (PBC). Applying the Jordan-Wigner transformation to the term $\hat{\sigma}_L^{+}\hat{
\sigma}^{-}_{L+1} \equiv \hat{\sigma}_L^{+}\hat{\sigma}_{L+1}^{-}$, we obtain $(-1)^{N} \hat{c}^{\dagger}_L \hat{c}_1$, where $N$ is the number of fermions. In terms of spins, $(-1)^N$ maps to the value of the parity operator $\mathcal{P} = \prod \hat{\sigma}_j^x$. Therefore, the Hilbert spaces splits in two subspaces: the one with even parity ($p = 0$) and, correspondingly, antiperiodic boundary conditions (ABC) and the one with odd parity ($p = 1$) and PBC~\cite{Mbeng_2024}.\\

Since a unit cell consists of two cites, we define the Fourier transformation as follows:
\be
\begin{split}
\hat{c}_{2j} &= \sqrt\frac{2}{L} \sum_{k \in \mathcal{K}} e^{ikj}\hat{b}_k\\
\hat{c}_{2j + 1} &= \sqrt\frac{2}{L} \sum_{k \in \mathcal{K}} e^{ikj}\hat{c}_k\\.
\end{split}
\ee
The set of wave vectors depends on the parity sector. Specifically, for the even sector we have
\be
\mathcal{K}_{p = 0} = \{k = \pm \frac{(2n - 1)\pi}{L'},~\text{with}~n = 1,..,\frac{L'}{2} \}
\ee
and for the odd sector 
\be
\mathcal{K}_{p = 1} = \{k = \pm \frac{2n\pi}{L'},~\text{with}~n = -\frac{L'}{2} + 1,.0,..\frac{L'}{2} \},
\ee
where $L'$ is the number of unit cells.\\

We introduce a following Nambu vector: $\mathbf{\Psi}_k^{\dagger} = (b_k^{\dagger},~c_k^{\dagger},~b_{-k},~c_{-k})^T$. After some algebra, we can write down the Hamiltonian in the momentum space:
\be
\hat{H}_{p = 0,1} = \sum_{k \in \mathcal{K}_{p = 0,1}}\mathbf{\Psi}_k^{\dagger} H_k \mathbf{\Psi}_k
\ee
with
\be
H_k = \begin{pmatrix}
    -C_k & D_k\\
    D_k^{\dagger} & C_k\\
\end{pmatrix},
\ee
where
\be
C_k = \begin{pmatrix}
    2h & J f(\gamma,-k)\\
     J f(\gamma, k) & 2h\\
\end{pmatrix}
\ee
with
\be
f(\gamma, k) = (1 + \gamma)  + (1 - \gamma) e^{ik}
\ee
and
\be
D_k = \begin{pmatrix}
    0 & Jg(\gamma,\theta,k)\\
     -Jg(\gamma,\theta,-k) & 0\\
\end{pmatrix}
\ee
with
\be
g(\gamma,\theta,k) = (1 - \gamma) e^{2i\theta}e^{-ik} - (1 + \gamma) e^{-2i\theta}.
\ee
Now, we can immediately obtain the excitation spectrum:
\be
\begin{split}
     &\varepsilon_k^{(\pm)}/J = \left[4 \left(\gamma ^2+(h/J)^2+1\right)-4 \left(\gamma ^2-1\right) \sin ^2(2 \theta ) \cos (k)\right.\\
     &\left. \pm \frac{1}{2}\left\{64 \left(\gamma ^2+(h/J)^2-\left(\gamma ^2-1\right) \sin ^2(2 \theta ) \cos (k)+1\right)^2\right.\right.\\
     &\left.\left.-8 \left(4 \left(\gamma ^2-1\right)^2 \cos (4 \theta )+\left(\gamma ^2-1\right)^2 \cos (8 \theta ) + 3 \left(\gamma ^2-1\right)^2\right.\right.\right.\\
     &+\left.\left.\left.8 (h/J)^4+16 \left(\gamma ^2-1\right) (h/J)^2 \cos ^2(2 \theta ) \cos (k)\right)
     \right\}^{1/2}\right]^{1/2}\\
\end{split}
\ee

For the simplest case of the TKSC without local inversion symmetry breaking, the expression above immediately simplifies to
\be
\begin{split}
    &\varepsilon_k^{(+/-)}/J = 2\left[[\cos{2k}\sin^2{2\theta} + (h/J)^2 + 1]  \right.\\
    &\pm \left. \sqrt{4(h/J)^2\cos^2{k} - \cos^4{2\theta} + (1 + \cos{2k}\sin^2{2\theta})^2}  \right]^{1/2}
\end{split}. \label{QP_spectrum}
\ee
One can easily find that the gap closes at
\be
h/J = \sqrt{1 - \gamma^2}|\cos{2\theta}|.
\ee
This is the quantum critical point. Additionally, it is worth noting that the gap always closes linearly at $k = 0$:
\be
\varepsilon_k/J \simeq ck~\text{for}~k \rightarrow 0~\text{with}~c = 2 (1 - \gamma^2)\cos^2{2\theta}. \label{low_energ_spectrum}
\ee
In the case of $\gamma = 0$, the presence of the local inversion symmetry manifests itself in band crossings at $k = \pm \pi/2$: $\eps_{k = \pm \pi/2}^{+} = \eps_{k = \pm \pi/2}^{-}$ for arbitrary $h/J$ and $\theta$. On the contrary, for any non-zero $\gamma$, the gap at $k= \pm \pi/2$ is observed.\\

\section{Zero-energy eigenvalues} \label{zero_modes}
To enforce the structure alluded in Eq.\eqref{eigenvectors}, we introduce the swap operator defined as follows:
\be
\hat{S}\begin{pmatrix}
    u\\
    v\\
\end{pmatrix} = \begin{pmatrix}
    v^*\\
    u^*\\
\end{pmatrix}.
\ee
This operator has two eigenvalues, $\lambda_+ = +1$ and $\lambda_- = -1$. The corresponding eigenvectors are
\be
\Vec{e}_+ = \begin{pmatrix}
    u + v^{*}\\
    v + u^{*}\\
\end{pmatrix}~\text{and}~\Vec{e}_- = \begin{pmatrix}
    u - v^{*}\\
    v - u^{*}\\
\end{pmatrix}.
\ee

Let us assume that $\Vec{e}_1$ and $\Vec{e}_2$ are the eigenvectors computed by a numerical routine. We need to find their linear combination that satisfies the structure in Eq.\eqref{eigenvectors}. To do this, we first find the matrix of operator $\hat{S}$ in the basis of these vectors. Since $\Vec{e}_1$ and $\Vec{e}_2$ constitute an eigenspace, it is correct to assume that the action of $\hat{S}$ on each of these vectors can be written as their linear combination:
\be
    \hat{S}\Vec{e}_i = \alpha_i \Vec{e}_1 + \beta_i \Vec{e}_2.
\ee
Thus, the matrix $\hat{S}$ takes in this basis the form
\be
\hat{S} = \begin{pmatrix}
    \alpha_1 & \beta_1\\
    \alpha_2 & \beta_2\\
\end{pmatrix}.
\ee
The coefficients $\alpha_i$ and $\beta_i$ can be found with the help of the following identities:
\be
\begin{split}
    \alpha_i &=  \frac{(Se_i,e_2)(e_1,e_2) - (Se_i, e1)}{(e_1,e_2)^2 - 1}\\
    \beta_i &= \frac{(Se_i,e_1)(e_1,e_2) - (Se_i, e2)}{(e_1,e_2)^2 - 1}.\\
\end{split}
\ee
The next step is to calculate the eigenvectors of $\hat{S}$, $\Vec{e}_+$ and $\Vec{e}_{-}$, and then combine them into $\frac{1}{2}\left(\Vec{e}_+ \pm \Vec{e}_{-} \right)$. The latter vectors yield the coefficients that we use to construct the desired linear combinations of $\Vec{e}_1$ and $\Vec{e}_2$.\\

\section{Finite-size effects} \label{finite_size}
\begin{figure}[h!]
\includegraphics[width=0.5\textwidth]{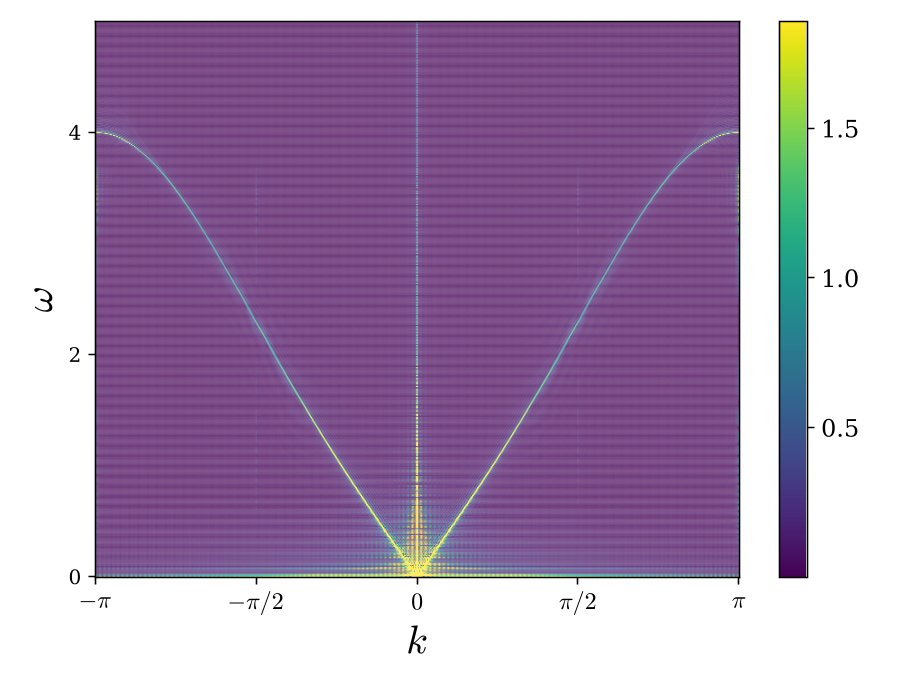}
\caption{The structure factor $S(k,\omega)$ for the TKSC. No cut-off is applied in the Fourier transformation. The vertical line at $k = 0$ indicates the finite size effect. $\theta = \pi/10$, $h/J = \cos{2\theta}$, $L = 500$.} \label{vert_line}
\end{figure}
\begin{figure}[h!]
\includegraphics[width=0.5\textwidth]{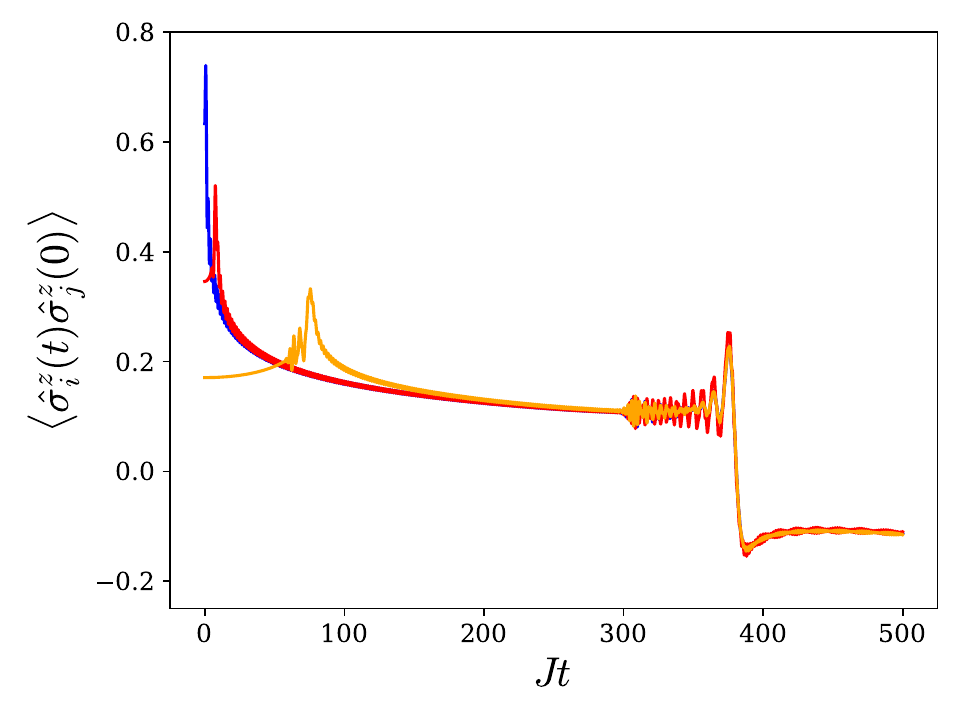}
\caption{Correlation functions $\langle \hat{\sigma}_{i}^z(t) \hat{\sigma}_{j}^z(0)\rangle$ for different $i$ and $j$ for the TKSC with $L = 500$ sites: $i = j = 250$ (the blue line); $i = 244$, $j = 254$ (the red line); $i = 200$, $j = 300$ (the orange line). $\theta = \pi/10$, $h/J = \cos{2\theta}$} \label{corrs_fin_size}
\end{figure}

Fig.\ref{vert_line} shows the most striking manifestation of the finite-size effect in our system, namely, the vertical line at $k = 0$ in the plot of the structure factor $S(k,\omega)$. To understand the origin of this feature, we turn out attention to the correlation function in real space (see Fig.\ref{corrs_fin_size}). The finite size effect is exhibited in the form of echos~\cite{PhysRevB.56.11659}, with the first echo occurring at the same time for all correlation functions, regardless the distance between the corresponding sites. This explains the emergence of the feature at $k = 0$.\\

To eliminate this artefact, we introduce a cut-off. Specifically, we multiply our data by factor $e^{-t^2/\tau^2}$. This smooth cut-off does not cause any new effects; and the parameter $\tau$ is chosen to be strong enough to eliminate manifestations of the finite-size effects. In our calculations, we set $\tau = T/4$, where $T = 500$ is the total measurement time.

\section{Ground state magnetization} \label{delta_peak}
In order to extinguish the delta peak discussed in the section \ref{numerics}, we subtract the ground state magnetization. Specifically, we compute connected correlation functions defined as
\be
\langle \langle \hat{\sigma}_{i}^z(t) \hat{\sigma}_{j}^z(0)\rangle \rangle = \langle \hat{\sigma}_{i}^z(t) \hat{\sigma}_{j}^z(0)\rangle - \langle \hat{\sigma}_{i}^z(t)\rangle\langle \hat{\sigma}_{j}^z(0)\rangle.
\ee

However, if we try to calculate $\langle \hat{\sigma}^z_i \rangle$ in the fermionic representation, we obtain zero, since $\hat{\sigma}^z_i$ in terms of fermions (see Eq.\eqref{JW_trafo}) does not conserve the number of fermions. The reason behind this lies in the two-fold degeneracy of the ground state in the thermodynamic limit~\cite{Mbeng_2024} and the fact that the physical ground states are linear combinations of the ground states in the PBC and ABC sectors (see Appendix \ref{eps_tkc}). \\

To understand this, we consider the simplest possible example of the Ising chain at zero magnetic field. This system has two physical ground states: $|\uparrow \uparrow ... \uparrow \rangle$ and $|\downarrow \downarrow ... \downarrow \rangle$. The ground states in the even parity ($p = 0$, APC) sector and the odd parity ($p = 1$, PBC) sectors are the following combinations of the above mentioned physical ground states:
\be
\begin{split}
    |GS\rangle_+ &= \frac{1}{\sqrt{2}}\left(|\uparrow \uparrow ... \uparrow \rangle +  |\downarrow \downarrow ... \downarrow \rangle \right),~~p = 0\\
    |GS\rangle_{-} &= \frac{1}{\sqrt{2}}\left(|\uparrow \uparrow ... \uparrow \rangle -  |\downarrow \downarrow ... \downarrow \rangle \right),~~p = 1.\\
\end{split}
\ee
Note that the action of the operator $\hat{\sigma}_i^z$ changes the parity sector. Indeed, $\hat{\sigma}_i^z |GS\rangle_{\pm} = |GS\rangle_{\mp}$. \\

Therefore, to measure the ground state magnetization, we introduce the operator $\hat{\Gamma}$ that changes the parity sector: $\hat{\Gamma} |GS\rangle_{\pm} = |GS\rangle_{\mp}$, and instead measure $\langle \hat{\Gamma}\hat{\sigma}_i^z \rangle$ in place pf $\langle \hat{\sigma}_i^z \rangle$. The operator $\hat{\Gamma}$ is defined as the sum of the zero mode creation and annihilation operators. Namely, if $\varepsilon_0 = 0$, the parity change operator is defined using the corresponding Bogoliubov operators
\be
\hat{\Gamma} = \hat{\eta}_0 + \hat{\eta}_0^{\dagger}.
\ee

Similar to the two-point correlation function, $\langle \hat{\sigma}_i^z \rangle$ can be written in terms of Pfaffians. We will not show the corresponding calculations in detail; however, for those interested, we provide the expressions for the corresponding matrix entries (the matrix is similar to Eq.\eqref{pfaff_matrix}; see also the definition of $\hat{A}$ and $\hat{B}$ in Eq.\eqref{AB_def}):
\be
\begin{split}
    \langle \hat{\Gamma}\hat{A}_i\rangle &= U_{i0} + V_{i0}\\
    \langle \hat{\Gamma}\hat{B}_i\rangle &= U_{i0} - V_{i0}.\\
\end{split}
\ee

\end{document}